\documentclass{article}
\usepackage{braket}
\usepackage{geometry}
\usepackage{graphicx}
\usepackage[hidelinks]{hyperref}
\usepackage{physics}
\usepackage{amsthm}
\usepackage{amsfonts}
\usepackage{xcolor}

\newtheorem{corollary}{Corollary}
\newtheorem{lemma}{Lemma}
\newtheorem{theorem}{Theorem}

\usepackage{changepage}
\newenvironment{widerequation}{%
    \begin{adjustwidth}{-1cm}{-1cm}\begin{equation}}
    {\end{equation}\end{adjustwidth}}

\begin{document}

\author{D. O'Malley$^{1*}$, J. M. Henderson$^2$, E. Pelofske$^2$, S. Greer$^{3}$, Y. Suba\c{s}\i$^2$, J. K. Golden$^2$, \\ R. Lowrie$^4$, S. Eidenbenz$^2$}
\date{\small $^1$Computational Earth Science (EES-16), Los Alamos National Laboratory\\
$*$omalled@lanl.gov\\
$^2$Information Sciences (CCS-3), Los Alamos National Laboratory\\
$^3$Massachusetts Institute of Technology\\
$^4$Computational Physics and Methods (CCS-2), Los Alamos National Laboratory}

\title{A near-term quantum algorithm for solving linear systems of equations based on the Woodbury identity}
\maketitle

\begin{abstract}
    Quantum algorithms for solving linear systems of equations have generated excitement because of the potential speed-up involved and the wide applicability of solving linear equations.
    However, applying these algorithms can be challenging.
    The Harrow-Hassidim-Lloyd algorithm and improvements thereof require complex subroutines, such as Hamiltonian simulation, that are suitable for fault-tolerant hardware but are ill-suited for contemporary noisy quantum computers.
    While variational algorithms are designed for today's machines, they involve expensive optimization loops, which can be prone to barren plateaus and local optima.
    We introduce a quantum algorithm for solving linear systems of equations that avoids these problems.
    Our algorithm is based on the Woodbury identity, which analytically describes the inverse of a matrix that is a low-rank modification of another easily-invertible matrix.
    This approach is thus applicable for systems that can be specified in the Woodbury identity's form, and it utilizes only basic quantum subroutines, such as the Hadamard test or the swap test, so it is well-suited to current hardware.
    Furthermore, there is no optimization loop, so barren plateaus and local optima do not present a problem, and the low-rank aspect of the identity enables efficient information transfer to and from the quantum computer.
    We demonstrate that this approach can produce accurate results on current hardware, including by estimating an inner product involving the solution of a system of more than 16 million equations with 2\% error using IBM's Auckland machine.
    To our knowledge, this is the largest system of equations to be solved with such accuracy on a quantum computer.
\end{abstract}

\section{Introduction}
Solving a linear system of equations is often a critical step in science and engineering problems.
For this reason, significant research has produced a variety of algorithms for solving linear equations.
Classical computers use iterative methods (such as the conjugate gradient method), direct methods (such as LU decomposition), or other methods that exploit the structure of a system of equations.
The structure of linear systems is reflected in  their matrix representations, and choosing the best algorithm often depends on that matrix structure.
For example, iterative methods are often used on large, sparse matrices, whereas direct methods are often used on small, dense matrices.

There are two primary classes of quantum algorithms for solving linear systems: algorithms that are guaranteed to succeed with high probability---often at the cost of subroutines that are difficult to implement---and variational algorithms, which are less sure to succeed, but more error-resistant. 
As with classical algorithms, matching the problem to the algorithm is crucial.
The first class was pioneered by the Harrow-Hasidim-Lloyd (HHL) algorithm \cite{harrow2009quantum}, with later improvements \cite{ambainis2012variable,childs2017quantum,costa2021optimal,jennings2023efficient,subasi2019quantum}.
In this paper we will use the HHL algorithm as a representative for all the algorithms in this first class.
Efficiently solving a system with the HHL algorithm requires that the system have specific properties \cite{aaronson2015read}.
These properties include that the equations be well-conditioned, that the matrix be efficiently simulatable (which holds, for example, if the system is sparse), that the right-hand side of the equation can be efficiently prepared, and that desirable information be efficiently extractable from the solution.
Additionally, HHL is built on a foundation of complex quantum subroutines such as Hamiltonian simulation, amplitude amplification, and quantum phase estimation.
These subroutines are suitable for fault tolerant devices, and therefore, HHL is not expected to be useful in the near term.
(Even algorithms that only require Hamiltonian simulation \cite{subasi2019quantum} will likely require error correction.)
However, these algorithms avoid the optimization issues of the second class of quantum algorithms for linear systems, to which we next turn.

Algorithms in the second class (i.e., variational algorithms) come in many varieties \cite{bravo2019variational,huang2019near,xu2021variational}, but have common characteristics. 
For example, the VQLS algorithm \cite{bravo2019variational} uses a variational optimization loop to find the solution, which involves challenges such as local optima and barren plateaus.
We will use the VQLS algorithm as a representative for all the algorithms in this second class.
Another challenge for VQLS is the requirement that the quantum computer must evaluate the optimization's loss function efficiently, thus constraining the systems that VQLS can solve efficiently.
Additionally, even for those problems that VQLS can solve, verifying that the optimization loop found the solution incurs additional computational cost \cite{somma2021complexity}.
Nonetheless, one advantage of variational approaches is the relative simplicity of the underlying quantum algorithms (e.g., the Hadamard test), which gives these approaches near-term potential.

Here, we describe a novel approach to solving systems of equations that combines some of the strengths of the HHL and VQLS approaches.
Like HHL, our approach does not require a variational optimization loop.
Therefore, it can reliably solve the system of equations, modulo sampling and hardware noise.
Like VQLS, our approach is built on a foundation of simple quantum algorithms like the Hadamard test or, in some cases, the swap test.
Therefore, it can be applied in the near term.
Our approach is based on the Woodbury matrix identity \cite{woodbury1950inverting,woodbury1949stability,guttman1946enlargement,hager1989updating}, which gives a formula for the inverse of a matrix that is a low-rank modification of an easily invertible matrix.
So, while it has some of the strengths of HHL and VQLS without some of the more serious drawbacks, it is limited to applications that can be specified with the Woodbury identity's form.

Fortunately, this admitted limitation includes several important applications \cite{hager1989updating}.
For example, it is used in uncertainty quantification as part of the Kalman filter \cite{mandel2006efficient}, in geophysical imaging \cite{lin2018efficient}, to improve deep generative flows in machine learning \cite{lu2020woodbury}, and in the oft-used Broyden-Fletcher-Goldfarb-Shanno optimization algorithm \cite{fletcher2013practical}, as well as variants like the limited memory Broyden-Fletcher-Goldfarb-Shanno algorithm \cite{byrd1995limited}.
The remainder of this manuscript discusses how algorithms based on the Woodbury matrix identity can be implemented on a quantum computer and tests the approach on quantum computing hardware.

\section{Algorithms based on the Woodbury matrix identity}
We want to solve a linear system of $N$ equations of the form
\begin{equation}
\label{eq:linsystem}
    (A + UCV)x = b\, ,
\end{equation}
where $A$ is an $N\times N$ matrix, $C$ is $k\times k$, $U$ is $N\times k$, and $V$ is $k\times N$. We assume that $b\equiv\ket{b}$ is a normalized vector that can be prepared efficiently on a quantum computer ($\ket{b}=B\ket{0}$ where $B$ is a unitary with cost, e.g., $polylog(N)$).
We assume that $(A+UCV)$ is nonsingular, and the form of equation \ref{eq:linsystem} allows us to use the Woodbury matrix identity to compute the matrix inverse:
\begin{equation}
    (A + UCV)^{-1} = A^{-1} - A^{-1}U(C^{-1}+VA^{-1}U)^{-1}VA^{-1} \, .
    \label{eq:woodbury}
\end{equation}
We explore using this identity to solve equation \ref{eq:linsystem} with quantum computers, starting with simple cases and working towards more complex cases.
When applying the Woodbury matrix identity, we assume that $A$ is easy to invert and $k$ is small compared to $N$.
Note that $(C^{-1}+VA^{-1}U)^{-1}$ is a $k$-by-$k$ matrix, so we may consider it small and manageable for a classical computer.
We assume
\begin{eqnarray}
    U &=& \sum_{i=0}^{k-1} \alpha_i \ket{u_i}\bra{i}\, , \\
    V &=& \sum_{i=0}^{k-1} \beta_i \ket{i}\bra{v_i}\, , \\
    \ket{u_i} &=& U_i \ket{0}\, , \\
    \ket{v_i} &=& V_i \ket{0}\, , \\
\end{eqnarray}
where $U_i$ and $V_i$ are unitaries that can be efficiently implemented on a quantum computer.
We may thus assume that quantities like $\braket{v_i}{u_j}$ can be computed using the Hadamard test, since $\braket{v_i}{u_j}=\expval{V_i^\dagger U_j}{v_i}$. This is possible in an access model where we can implement controlled versions of these unitaries.
It is worth noting that we need not make further assumptions about the properties that sets of states $\{\ket{u_i}\}_i$ and $\{\ket{v_i}\}_i$ have; for example, they need not be mutually orthogonal.
In cases where the phase of the inner product is known (e.g., if the inner product is known \emph{a priori} to be positive and real), these inner products can also be computed using the swap test \cite{buhrman2001quantum}.
The swap test does not require that the unitaries be controlled, so these cases simplify the circuits needed to implement the unitaries and require a weaker access model.

Computing inner products is at the heart of using the Woodbury matrix identity on a quantum computer; the key to any quantum speed-up for solving linear systems of equations with this approach is that the quantum computer must compute the inner products faster than a classical computer.
While we do not consider computation of all possible inner-products in this paper, section \ref{sec:woodbury_speedup} proves that the Woodbury approach provides an exponential speedup over classical methods when computing inner products that can be expressed as Forrelation problems \cite{aaronson2015forrelation}.

\subsection{U and V are rank-1, A=I, C=I}\label{sec:simplest}
Assuming that $A=I$, $C=I$ and $U$ and $V$ are rank-1 ($k=1$), the Woodbury identity becomes
\begin{equation}
    (I + UV)^{-1} = I - \frac{UV}{1 + VU}  \label{eq:case1}
\end{equation}
and the low rank matrices reduce to vectors: $U=\alpha_0 \ket{u_0}$ and $V=\beta_0 \bra{v_0}$ (or, they may be thought of as an $n$-by-1 matrix and a 1-by-$n$ matrix).
We can then describe the solution to equation \ref{eq:linsystem}  as
\begin{eqnarray}\nonumber
    \ket{x} &=& (I + UV)^{-1}\ket{b} \\ \nonumber
     &=& \left(I - \frac{UV}{1 + VU}\right) \ket{b}\\\nonumber
    &=& \left(I - \frac{\alpha_0\beta_0\ket{u_0}\bra{v_0}}{1 + \alpha_0\beta_0\braket{v_0}{u_0}}\right)\ket{b} \\
    &=& \ket{b} - \frac{\alpha_0\beta_0\braket{v_0}{b}}{1 + \alpha_0\beta_0\braket{v_0}{u_0}}\ket{u_0} \, .
\end{eqnarray}
Note that $\ket{x}$ in general is not a normalized quantum state. 
Let $\ket{z}=Z\ket{0}$, where $Z$ is a unitary that can be implemented efficiently.
Then,
\begin{equation}
    \braket{z}{x} = \braket{z}{b} - \frac{\alpha_0\beta_0\braket{v_0}{b}}{1 + \alpha_0\beta_0\braket{v_0}{u_0}}\braket{z}{u_0}
    \label{eq:case1final}
\end{equation}
and $\braket{z}{x}$ can be estimated by computing the four inner products $\braket{z}{b}$, $\braket{v_0}{b}$, $\braket{v_0}{u_0}$, and $\braket{z}{u_0}$.
As mentioned in the previous section, these inner products can be estimated using the Hadamard test.

\subsubsection{The Woodbury approach for computing expectations}
Often, the goal of a quantum algorithm is to estimate the expectation value of a Hermitian operator, $O$.
Under the same constraints of $A=I$, $C=I$ and $U$ and $V$ are rank-1 ($k=1$), the Woodbury approach for computing $\bra{x}O\ket{x}$ gives
\begin{eqnarray*}
\bra{x}O\ket{x} &=& \left[ \bra{b} - \left(\frac{\alpha_0\beta_0 \braket{v_0}{b}}{1 + \alpha_0\beta_0\braket{v_0}{u_0}}\right)^* \bra{u_0} \right] O \left[ \ket{b} - \left(\frac{\alpha_0\beta_0 \braket{v_0}{b}}{1 + \alpha_0\beta_0\braket{v_0}{u_0}}\right) \ket{u_0} \right] \\
&=& \bra{b}O\ket{b} - \left(\frac{\alpha_0\beta_0 \braket{v_0}{b}}{1 + \alpha_0\beta_0\braket{v_0}{u_0}}\right) \bra{b}O\ket{u_0} \\
&~& - \left(\frac{\alpha_0\beta_0 \braket{v_0}{b}}{1 + \alpha_0\beta_0\braket{v_0}{u_0}}\right)^* \bra{u_0}O\ket{b} + \left|\frac{\alpha_0\beta_0 \braket{v_0}{b}}{1 + \alpha_0\beta_0\braket{v_0}{u_0}}\right|^2\bra{u_0}O\ket{u_0} \\
&=& \bra{b}O\ket{b} - 2Re\left(\frac{\alpha_0\beta_0 \braket{v_0}{b}}{1 + \alpha_0\beta_0\braket{v_0}{u_0}}\right) \bra{b}O\ket{u_0} \\
&~&  + \left|\frac{\alpha_0\beta_0 \braket{v_0}{b}}{1 + \alpha_0\beta_0\braket{v_0}{u_0}}\right|^2\bra{u_0}O\ket{u_0}\, .
\end{eqnarray*}
(In the last step, we apply the fact that $\bra{b}O\ket{u_0}^*=\bra{u_0}O^\dagger\ket{b}=\bra{u_0}O\ket{b}$ since $O$ is Hermitian.)
The resulting expression is analogous to equation \ref{eq:case1final}, in the sense that obtaining the left-hand-side value requires computing the inner products $\braket{v_0}{b}$ and $\braket{v_0}{u_0}$, as well as the quantities $\bra{b}O\ket{b}$, $\bra{u_0}O\ket{u_0}$, and $\bra{b}O\ket{u_0}$.
The two inner products are readily computed with the Hadamard test, and $\bra{b}O\ket{b}$ and $\bra{u_0}O\ket{u_0}$ are also obtainable using the Hadamard test.
Therefore, we need only consider how to compute $\bra{b}O\ket{u_0}$.
If we assume that $O=\sum_i \xi_i W_i$ is a linear combination of unitaries, then $\bra{b}O\ket{u_0}$ can be estimated by observing that
\begin{equation}
    \bra{b}O\ket{u_0} = \bra{0}U_b^\dagger\sum_i \xi_i W_i U_0\ket{0} = \sum_i \xi_i \bra{0}U_b^\dagger W_i U_0\ket{0}
    \label{eq:bou0}
\end{equation}
and using the Hadamard test to estimate each expectation value on the right hand side of equation \ref{eq:bou0}.

\subsubsection{Error and complexity analysis}
We next consider the relationship between solution error and the complexity of the Woodbury approach when $A=I$, $C=I$ and $U$ and $V$ are rank-1 ($k=1$).
First, we determine the number of shots required to compute $\braket{z}{x}$ (equation \ref{eq:case1final}) with a precision of $O(\epsilon)$.
To do so, we determine the number of shots needed to compute each of the inner products that contribute to $\braket{z}{x}$ ($N_{\braket{z}{b}}$, $N_{\braket{v_0}{b}}$, $N_{\braket{v_0}{u_0}}$, and $N_{\braket{z}{u_0}}$) such that the overall error of $\braket{z}{x}$ is $O(\epsilon)$.
By expanding the right-hand side of equation \ref{eq:case1final} to first order with respect to $\braket{z}{b}$, $\braket{v_0}{b}$, $\braket{v_0}{u_0}$, and $\braket{z}{u_0}$, and by applying the relationship between shots required and error for the Hadamard test, we can show that $N_{\braket{z}{b}}$, $N_{\braket{v_0}{b}}$, $N_{\braket{v_0}{u_0}}$, and $N_{\braket{z}{u_0}}$ are given by
\begin{eqnarray}
N_{\braket{z}{b}} &=&  \left[ \frac{1}{\epsilon}\right]^2\, , \label{eq:n1} \\
N_{\braket{v_0}{b}} &=&  \left[\frac{\alpha_0\beta_0\braket{z}{u_0}}{\epsilon(1 + \alpha_0\beta_0 \braket{v_0}{u_0})}\right]^2\, , \label{eq:n2} \\
N_{\braket{v_0}{u_0}} &=& \left[ \frac{\alpha_0^2\beta_0^2 \braket{z}{u_0}\braket{v_0}{b}}{\epsilon (1+\alpha_0\beta_0\braket{v_0}{u_0})^2} \right]^2\, , \label{eq:n3} \\
N_{\braket{z}{u_0}} &=& \left[ \frac{ \alpha_0 \beta_0 \braket{v_0}{b}}{\epsilon (1 + \alpha_0\beta_0\braket{v_0}{u_0})} \right]^2\, . \label{eq:n4}
\end{eqnarray}
Therefore, the optimal number of shots for each inner product estimation depends on the true value of the inner products. 
While this presents an interesting challenge for optimal sampling methods, we do not employ an adaptive sampling strategy in the results section; instead, for convenience, we use a fixed number of shots for each inner product.

It is worth noting that the quantity $\gamma= \alpha_0 \beta_0/(1+\alpha_0\beta_0\braket{v_0}{u_0})$ appears in equations \ref{eq:n2}-\ref{eq:n4}. 
When $(I+UV)$ is Hermitian, it can be shown that the matrix $(I+UV)$ has only one eigenvalue different from 1 given by $(1+\alpha_0\beta_0 \braket{v_0}{u_0})$. 
In the limit of large condition number, it can be shown that $\gamma=O(1/\lambda_s)$, where $\lambda_s$ is the smallest eigenvalue (in magnitude) of $(I+UV)$.
Thus, the sampling complexity scales as $O(1/(\lambda_s^2 \epsilon^2))$.

Most quantum algorithms in the literature~\cite{ambainis2012variable,childs2017quantum,harrow2009quantum} output the normalized state $\ket{x_\text{norm}}=\ket{x}/\|\ket{x}\|$. 
Computing $\braket{z}{x}$ requires information about the norm of the solution vector. 
A naive approach is thus to estimate $\braket{z}{x_\text{norm}}$ and $\|\ket{x}\|$ separately. 
However, $\braket{z}{x}$ can be estimated more directly as follows. 
On the way to preparing the normalized state $\ket{x_\text{norm}}$, the HHL algorithm makes use of a unitary that block-encodes the matrix inverse:
\begin{align}
    \label{eq:generalQLSA}
    U_{A^{-1}} \ket{0}_\text{anc}\otimes \ket{b} \approx \ket{0}_\text{anc} \otimes \frac{A^{-1}}{a} \ket{b} + \ket{0^\perp} 
\end{align}
where $\ket{0^\perp}$ is orthogonal to the state $\ket{0}_\text{anc}$ and $a\ge 1/\lambda_s$ is a constant necessary to ensure that the right hand side is a normalized state.
We observe that
\begin{align}
    \frac{1}{a} \bra{z}A^{-1}\ket{b} &\approx (\bra{0}_\text{anc} \otimes \bra{z}) U_{A^{-1}} (\ket{0}_\text{anc}\otimes\ket{b}) \\ 
    &= (\bra{0}_\text{anc} \otimes \bra{z}) U_{A^{-1}}(I\otimes U_b U_z^\dagger)    (\ket{0}_\text{anc} \otimes \ket{z}) \; .
\end{align}
We can compute this quantity using the Hadamard test as in the Woodbury algorithm described above.
There are two sources of error.
First, there is systematic error due to the fact that the block-encoding is not exact. 
Then there is statistical error due to finite number of samples used in Hadamard test. 
If we solely consider the statistical error, we see that the number of samples required to obtain an estimate to $\bra{z} A^{-1}\ket{b}$ with error $\epsilon$ scales as $O(1/(\lambda_s^2 \epsilon^2))$.
This matches the complexity of the Hadamard test for the Woodbury approach.
Reducing the systematic error can be costly for the original HHL algorithm because of its poor scaling with respect to error, but there are other approaches based on Eq.~\eqref{eq:generalQLSA} with exponentially improved scaling with error~\cite{childs2017quantum}, and for those approaches the systematic error can be made small without significant increase in complexity.

We next consider the complexity in terms of the condition number of the matrix $I + UV$, because, generally, the complexity of quantum algorithms for linear systems is expressed in terms of the condition number $\kappa$ of the matrix in the linear system \cite{ambainis2012variable,bravo2019variational,childs2017quantum,costa2021optimal,harrow2009quantum,subasi2019quantum}. 
This decision to express error in terms of $\kappa$ results from the fact that most quantum algorithms prepare the normalized quantum state $\ket{x}/\|\ket{x}\|$ and give guarantees on \textit{additive error}, or error that does not depend upon the true value of $\ket{x}$.
(Contrast this with equations \ref{eq:n1} through \ref{eq:n4}, which provide the number of shots required to achieve a certain error \textit{in terms of} the true values being estimated.)
Full analysis is included in appendix \ref{app:error-analysis}, and here we summarize the results.
In most cases of interest, the Woodbury approach recovers the $O(\kappa^2)$ scaling of HHL, with several notable exceptions.
In particular, for most instances where $I+UV$ is Hermitian or $\alpha_0\beta_0 = 1$ the complexity of the Woodbury approach does not diverge as $\kappa \to \infty$.
The difference comes from the fact that the form of equation \ref{eq:case1final} can be used to exploit additional structure in the linear system to be solved, as opposed to the one-size-fits-all approach of HHL.
As is discussed further in appendix \ref{app:error-analysis}, there are specific niche cases, e.g., $U = V = b$, when the Woodbury approach scales as $O(\kappa^4)$ (this is a ``trivial case'' where $\ket{x}=\ket{b}$), but in the general case, the dominant error comes from the fact that 
\begin{equation}
    \ket{\tilde{x}} \approx \text{sgn}(1+\alpha_0\beta_0\braket{v_0}{u_0})\ket{u},
\end{equation}
and
\begin{equation}
    \braket{v_0}{u_0} \approx -\frac{1}{\alpha_0\beta_0} +\frac{\alpha_0\beta_0}{\kappa}.
\end{equation}
Therefore, one must estimate $\braket{v_0}{u_0}$ to increasingly high levels of accuracy in order to accurately determine $\text{sgn}(1+\alpha_0\beta_0\braket{v_0}{u_0})$.
Since the necessary accuracy scales as $O(1/\kappa)$, and the accuracy of the Hadamard test scales as $O(1/\epsilon^2)$, the overall complexity in the general case (i.e., avoiding the aforementioned niche cases) scales as $O(\kappa^2/\epsilon^2)$.

Ref. \cite{arrasmith2020operator} describes strategies for optimizing the number of shots to achieve a desired precision when computing expectations using a Hermitian operator, $O$, as described above. As in the case of estimating an overlap, the condition number of the matrix influences the required number of shots.

\subsubsection{Proving an exponential speedup}\label{sec:woodbury_speedup}
We now prove that for problems which can be expressed as Forrelation problems, the Woodbury approach for linear systems offers an exponential speedup over classical methods.
Forrelation problems are defined in Ref. \cite{aaronson2015forrelation} as problems for which quantum computing provides a ``maximum'' amount of speedup, and they amount to determining whether ``one Boolean function is highly correlated with the Fourier transform of a second function'' \cite{aaronson2015forrelation}.
For inner-product computation problems that can be expressed in this way, we first prove a bound on the complexity of computing $\braket{z}{x}$ classically, and then show how that leads to an exponential speedup when applying the Woodbury approach.

\begin{theorem}
    Let $f,g : \{ 0, 1\}^n\rightarrow\{-1, 1\}$ and let $U_f$ and $U_g$ be unitary oracles for $f$ and $g$ respectively.
    Then any classical algorithm for evaluating $\braket{z}{x}$ within, say, $\epsilon=10^{-3}$ with bounded probability of error must make $\Omega\left(2^{n/2}/n\right)$ queries, where
    $\ket{x}$ is the unnormalized solution to $(I+UV)x=b$,
    $\ket{z}=U_f^{\dagger}H^{\otimes n}\ket{0}$,
    $\ket{b}=H^{\otimes n}U_gH^{\otimes n}\ket{0}$,
    $U=\frac{1}{2}H^{\otimes n}U_gH^{\otimes n}\ket{0}$, and
    $V=\bra{0}H^{\otimes n}U_f$.
    \label{thm:classical_polynomial}
\end{theorem}
\begin{proof}
    In the notation of the main text, we may write the vectors comprising the inner products as
    \begin{eqnarray*}
        \ket{v_0} &=& U_f^{\dagger}H^{\otimes n}\ket{0} \\
        \ket{u_0} &=& H^{\otimes n}U_gH^{\otimes n}\ket{0} \\
        \alpha_0\beta_0 &=& \frac{1}{2}
    \end{eqnarray*}
    Let $F = \bra{0}H^{\otimes n} U_f H^{\otimes n}U_gH^{\otimes n} \ket{0}$ be the Forrelation of $f$ and $g$ \cite{aaronson2015forrelation}.
    Plugging into equation \ref{eq:case1final} using the above expressions for the inner products and the definition of Forrelation $F$, we may express the inner product $\braket{z}{x}$ as a sum of Forrelations:
    \begin{eqnarray*}
        \braket{z}{x}
        = \braket{z}{b} - \frac{\alpha_0\beta_0\braket{v_0}{b}\braket{z}{u_0}}{1 + \alpha_0\beta_0\braket{v_0}{u_0}}
        = F - \frac{F^2}{2+F} = \frac{2F}{2 + F}
    \end{eqnarray*}
    Solving this equation for $F$ gives
    \begin{equation}
        F = \frac{2\braket{z}{x}}{2 - \braket{z}{x}}
        \label{eq:forrelation_linearsystems_connection}
    \end{equation}
    Suppose a classical algorithm could estimate $\braket{z}{x}$ within $\epsilon=10^{-3}$ with bounded probability of error using fewer than $\Omega\left(2^{n/2}/n\right)$ queries.
    Then equation \ref{eq:forrelation_linearsystems_connection} implies that the Forrelation problem could also be solved using fewer than $\Omega\left(2^{n/2}/n\right)$ queries.
    However, it has been shown that any classical algorithm for solving the Forrelation problem requires $\Omega\left(2^{n/2}/n\right)$ queries \cite{aaronson2015forrelation}.
    This contradicts our supposition that a classical algorithm could estimate $\braket{z}{x}$ within $\epsilon=10^{-3}$ with bounded probability of error using fewer than $\Omega\left(2^{n/2}/n\right)$ queries.
    Therefore, any classical algorithm to estimate $\braket{z}{x}$ within $\epsilon=10^{-3}$ with bounded probability of error requires $\Omega\left(2^{n/2}/n\right)$ queries.
\end{proof}

\begin{corollary}
    If $U_f$ and $U_g$ can be controlled, the Woodbury approach provides an exponential speed-up over classical algorithms.
\end{corollary}
\begin{proof}
    If $U_f$ and $U_g$ can be controlled, $F=\bra{0}H^{\otimes n} U_f H^{\otimes n}U_gH^{\otimes n} \ket{0}$ can be estimated by using the Hadamard test.
    If $\braket{z}{x}=C(F) = \frac{2F}{2 + F}$, then $C(F+h)\approx C(F) + \frac{4h}{(2+F)^2}$.
    This implies that estimating $\braket{z}{x}=C(F)$ to accuracy $\epsilon$ with bounded probability of error requires a number of shots given by
    \begin{equation}
        N_F=\left[ \frac{4}{\epsilon (2 + F)^2}\right]^2
    \end{equation}
    Since, by definition, $|F|\leq1$, we conclude that $N_F \leq 16/\epsilon^2$.
    This implies that for a fixed $\epsilon$ (say, $\epsilon=10^{-3}$), the cost of estimating $\braket{z}{x}$ to accuracy $\epsilon$ requires a fixed number of controlled queries.
    Theorem \ref{thm:classical_polynomial} allows us to conclude that the Woodbury approach provides an exponential speed-up compared to any classical algorithm.
\end{proof}

Note that $U_f$ and $U_g$ can be controlled; for example, when they are provided as quantum circuits, those unitary operations can be controlled.
Consequently, this result applies to oracles that are implemented on a programmable quantum computer.

\subsection{U and V are rank-k, A=I}\label{sec:a_is_i}
Assuming $A=I$, the Woodbury identity reduces to
\begin{equation}
    (I + UCV)^{-1} = I - U(C^{-1}+VU)^{-1}V\, .
\end{equation}
In this case, we can then describe the solution of (\ref{eq:linsystem}) as
\begin{eqnarray*}
    \ket{x} &=& (I + UCV)^{-1}\ket{b}\\
    &=& (I - U(C^{-1}+VU)^{-1}V)\ket{b} \\
    &=& \left[I - \left(\sum_i \alpha_i \ket{u_i}\bra{i}\right)\left(C^{-1}+\left\{\sum_{i,j}\alpha_i\beta_j\braket{v_j}{u_i}\ket{j}\bra{i}\right\}\right)^{-1}\left(\sum_j \beta_j \ket{j}\bra{v_j}\right)\right]\ket{b} \\
    &=& \ket{b} - \left(\sum_i \alpha_i \ket{u_i}\bra{i}\right)\left(C^{-1}+\left\{\sum_{i,j}\alpha_i\beta_j\braket{v_j}{u_i}\ket{j}\bra{i}\right\}\right)^{-1}\left(\sum_j \beta_j \ket{j}\braket{v_j}{b}\right)\, .
\end{eqnarray*}
Note again that $\ket{x}$ in general is not a normalized quantum state. 
The solution here is a linear combination of quantum states rather than a single quantum state, and is thus similar to results obtained using some variational approaches \cite{huang2019near}.
Taking an inner product of $\ket{x}$ with $\ket{z}$, we get
\begin{equation} \label{eq:16}
    \braket{z}{x} = \braket{z}{b} - \left(\sum_i \alpha_i \braket{z}{u_i}\bra{i}\right)\left(C^{-1}+\left\{\sum_{i,j}\alpha_i\beta_j\braket{v_j}{u_i}\ket{j}\bra{i}\right\}\right)^{-1}\left(\sum_j \beta_j \ket{j}\braket{v_j}{b}\right)\, .
\end{equation}
This requires the computation of $k^2 + 2k + 1$ inner products: $k^2$ of the form $\braket{v_j}{u_i}$, $k$ of the form $\braket{z}{u_i}$, $k$ of the form $\braket{v_j}{b}$, and the one $\braket{z}{b}$.
An algorithm for computing $\braket{z}{x}$ that uses a combination of classical and quantum resources proceeds as follows:
\begin{enumerate}
    \item \verb+y1=+$\braket{z}{b}$
    \item \verb+for i=1:k y2[i]=+$\alpha_i\braket{z}{u_i}$\verb+ end+
    \item \verb+for j=1:k y3[j]=+$\beta_j\braket{v_j}{b}$\verb+ end+
    \item \verb+for i=1:k for j=1:k y4[i,j]=+$\alpha_i\beta_j\braket{v_j}{u_i}$\verb+ end end+
    \item \verb|M=inv(inv(C)+y4)|
    \item \verb|return y1-transpose(y2)*M*y3|
\end{enumerate}
All steps in the algorithm are done classically, except for the inner products, which are computed using the quantum computer.
Note that the matrix inversion (\verb|inv(C)+y4|) requires $O(k^3)$ operations in general.
The scaling in $N$ will depend on the complexity of the unitaries required to prepare $\ket{z}$, $\ket{u_i}$, $\ket{v_j}$, and $\ket{b}$, so the complexity in $N$ is problem-dependent. It is also worth noting that this algorithm does not require embedding non-Hermitian matrices in a higher-dimensional space, unlike other algorithms such as HHL.
 
The error analysis for the case of Sec. \ref{sec:simplest} is quite involved, and as the analysis for determining the relationship between number of shots and precision is even more complex in this case, we do not work through it here.
Instead, we make the following observation: the condition number of the matrix to be inverted still plays a crucial role.
Specifically, the matrix inversion on the right-hand side of equation \ref{eq:16} is performed on a classical computer.
And because the entries of that matrix are obtained using the quantum algorithm, the inversion can be sensitive to the quantum algorithm's performance, requiring that each matrix element be obtained with high precision.
Thus the dependence of the number of shots necessary on the condition number is expected to be similar to that in the previous section.

\subsection{U and V are rank-k, A is unitary}
Assuming $A=Q$ is a unitary matrix, the Woodbury identity reduces to
\begin{equation}
    (Q + UCV)^{-1} = Q^\dagger - Q^\dagger U(C^{-1}+VQ^\dagger U)^{-1}VQ^\dagger\, .
\end{equation}
We can then describe the solution to $(Q + UCV)x=b$ as
\begin{eqnarray*}
    \ket{x} &=& (Q + UCV)^{-1}\ket{b}\\
    &=& (Q^\dagger - Q^\dagger U(C^{-1}+VQ^\dagger U)^{-1}VQ^\dagger)\ket{b} \\
    &=& \left[Q^\dagger - Q^\dagger\left(\sum_i \alpha_i \ket{u_i}\bra{i}\right)\left(C^{-1}+\left\{\sum_{i,j}\alpha_i\beta_j\bra{v_j}Q^\dagger\ket{u_i}\ket{j}\bra{i}\right\}\right)^{-1}\left(\sum_j \beta_j \ket{j}\bra{v_j}\right)Q^\dagger\right]\ket{b} \\
    &=& Q^\dagger\ket{b} - \left(\sum_i \alpha_i Q^\dagger\ket{u_i}\bra{i}\right)\left(C^{-1}+\left\{\sum_{i,j}\alpha_i\beta_j\bra{v_j}Q^\dagger\ket{u_i}\ket{j}\bra{i}\right\}\right)^{-1}\left(\sum_j \beta_j \ket{j}\bra{v_j}Q^\dagger\ket{b}\right)\, .
\end{eqnarray*}
As mentioned previously, $\ket{x}$ in general is not a normalized quantum state. 
Introducing $\ket{z}$ on the left-hand side, we get
\begin{equation}
    \braket{z}{x} = \bra{z}Q^\dagger\ket{b} - \left(\sum_i \alpha_i \bra{z}Q^\dagger\ket{u_i}\bra{i}\right)\left(C^{-1}+\left\{\sum_{i,j}\alpha_i\beta_j\bra{v_j}Q^\dagger\ket{u_i}\ket{j}\bra{i}\right\}\right)^{-1}\left(\sum_j \beta_j \ket{j}\bra{v_j}Q^\dagger\ket{b}\right)\, .
\end{equation}
The algorithm now proceeds essentially as described in the previous section.
Specifically, we must compute $k^2 + 2k + 1$ matrix multiplications/dot product combinations: $k^2$ of the form $\bra{v_i}Q^\dagger\ket{u_i}$, $k$ of the form $\bra{z}Q^\dagger\ket{u_i}$, $k$ of the form $\bra{v_j}Q^\dagger\ket{b}$, and the one $\bra{z}Q^\dagger\ket{b}$.
The Hadamard tests for these values are now more complicated than in the previous subsections because they all involve a $Q^\dagger$, but all values remain computable; whereas in earlier sections, we computed $\braket{v_j}{u_i}=\expval{U_iV_j^\dagger}{v_j}$ using the Hadamard test, we can now compute $\bra{v_j}Q^\dagger\ket{u_i}=\expval{Q^\dagger U_iV_j^\dagger}{v_j}$ using the same.
The other components ($\bra{z}Q^\dagger\ket{b}$, $\bra{z}Q^\dagger\ket{u_i}$, and $\bra{v_j}Q^\dagger\ket{b}$) are computed similarly.
Consequently, an algorithm for computing $\braket{z}{x}$ with a combination of classical and quantum resources proceeds analogously to that in section \ref{sec:a_is_i}, but with the right-hand side of each \texttt{yi}-value adjusted to the quantities described here.
So, the cost of the algorithm is increased only by the inclusion of the extra controlled unitary ($Q^\dagger$) when implementing the Hadamard test.

\section{Results}
We ran two sets of experiments---both simple, concrete examples of applying the method of section \ref{sec:simplest}---to assess the Woodbury approach when implemented on existing quantum hardware.
The first experiments used problems with increasingly-large sizes and uniform condition numbers.
These problems were solved using an IBM superconducting qubit quantum computer, and specifically, the twenty-seven qubit machine \texttt{ibm\_auckland} which has a sparse hardware graph called heavy-hex \cite{Chamberland_2020}. 
We then considered problems that could be expressed as Forrelation problems, such that they experience the provable speedup of section \ref{sec:woodbury_speedup}.
We ran these problems using Quantinuum's trapped-ion \texttt{H1-2} hardware with twelve qubits.

\subsection{Experiments with \texttt{ibm\_auckland}}\label{sec:results_ibm}
For the first set of problems, we set $U_0$, $V_0$, $Z$, and $B$ to be tensor products of Hadamard operators, and we set $\alpha_0=\beta_0=1$.
Therefore, $\ket{z}$, $\ket{b}$, $\ket{u_0}$, and $\ket{v_0}$ are all the uniform superposition ($\sum_{i=0}^{N-1} \ket{i}/\sqrt{N}$).
This renders all the inner products, $\braket{z}{b}$, $\braket{v_0}{b}$, $\braket{v_0}{u_0}$, and $\braket{z}{u_0}$, on the right-hand side of equation \ref{eq:case1final} equal to 1, so $\braket{z}{x} = 1 / 2$.
Having a problem that can be solved analytically makes for straightforward evaluation of the quantum computer's result quality.

\begin{figure}
\begin{centering}
    \includegraphics[width=0.7\textwidth]{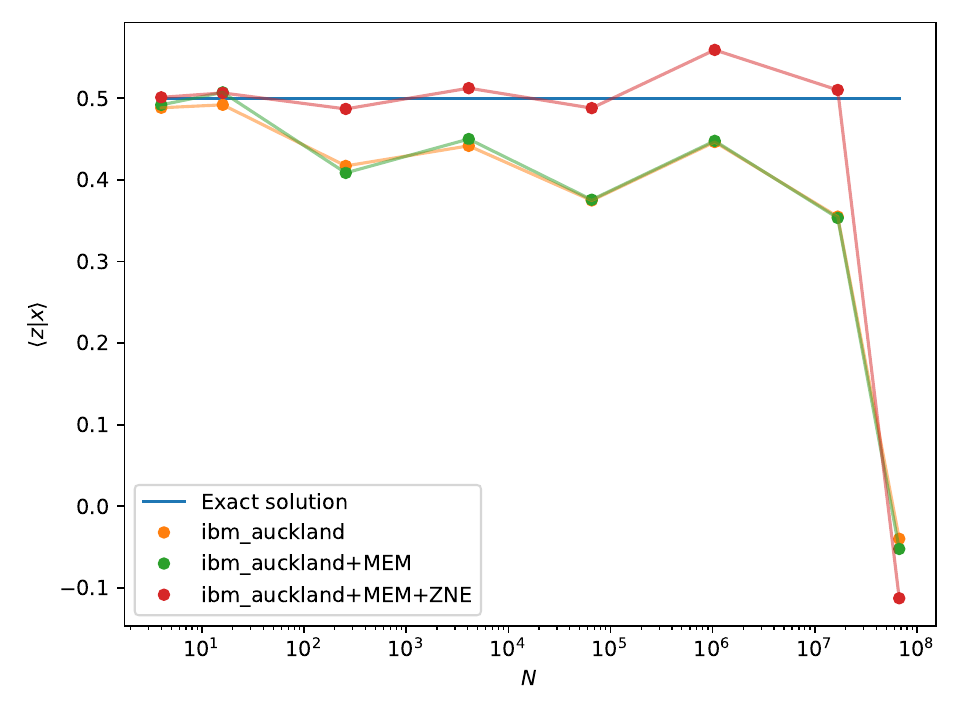}
    \caption{Estimates of $\braket{z}{x}$ for problems ranging in size from four equations/unknowns to $2^{26}$ equations/unknowns using our algorithm on 
    IBM Auckland hardware. The orange curve shows results without post-processing. The green curve shows results using measurement error mitigation (MEM). The red curve shows the results using measurement error mitigation and zero noise extrapolation (MEM+ZNE).
    }
    \label{fig:ibm_results}
\end{centering}
\end{figure}

Using \texttt{ibm\_auckland}, we studied this example with two post-processing methods to mitigate noise.
The following three points apply to all the results of Figure \ref{fig:ibm_results}.
First, we did not perform manual circuit optimization; although for this simple problem, the circuits could be heavily optimized by hand, we rely on Qiskit's transpiler to perform all optimizations (by setting the transpiler's optimization setting to its highest value of three).
Because the two qubit connectivity graph of \texttt{ibm\_auckland} is sparse, SWAP gates must be used to route qubits around the hardware graph so that the logical circuit can be implemented. 
Second, we took $10^5$ shots for estimating each inner product.
Although the optimal number of shots depends upon the value of the true solution, we chose a fixed shot count, because the goal of these experiments was to assess the accuracy of the Woodbury method absent considerations about an adaptive sampling method for determining optimal shot count.
Third, we used the Hadamard test; although we knew the phase of each inner product in these problems, we opted to apply the more-general Hadamard test (and not the swap test), since the former is applicable in more situations than the latter.
In so doing, we exploited the fact that we know each inner product is real; this allows us to use solely one Hadamard test per inner product, while inner products that may be complex would require two Hadamard tests.
Leveraging the fact that the inner products are real is applicable to a broad set of cases: any in which the matrices ($A$, $C$, $U$, and $V$) and vectors ($z$ and $b$) are real-valued.

We computed the inner products using three levels of post-processing: first, we used no post-processing; the output from the Hadamard tests directly estimated the inner products, which were then plugged into equation \ref{eq:case1final}.
The second method applied linear measurement error mitigation (MEM) by solving a $2$-by-$2$ system of equations to correct the estimate of each inner product \cite{qiskit}.
Third and finally, we combined measurement error mitigation with zero noise extrapolation using linear extrapolation \cite{temme2017error}.
Details on the use of zero noise extrapolation are given in appendix \ref{app:zne}.

Figure \ref{fig:ibm_results} illustrates the results for systems with between four and $2^{26}$ equations, or $4\leq N \leq 2^{26}$.
All three post-processing approaches are reasonably accurate for problems up to $N=2^{24}$, but at $N=2^{24} (\approx 10^7)$, the estimates with no post-processing and measurement error mitigation both have $29\%$ relative error, while the estimate with measurement error mitigation and zero noise extrapolation is reduced to $2\%$.
When $N=2^{26}$, there is a sharp decrease in accuracy, and the reasons for this are not clear. Two hypotheses are that the larger circuit was forced to use a part of the chip with higher error rates and/or that the compiler struggled to optimize the larger circuit.

Additionally, for all problems, measurement error mitigation provides at best a slight improvement in accuracy compared to the solutions with no post-processing.
Conversely, zero noise extrapolation significantly reduces the error in the estimate of $\braket{z}{x}$.
It should be noted that for a random state, $\ket{y}$, $\braket{z}{y}\approx 0$, or more precisely $E[|\braket{z}{y}|^2]\sim 1/N$.
These results suggest that the dominant source of error is hardware noise during circuit execution, which has a mild dependence on the problem size due to the simplicity of the quantum circuits involved.

\subsection{Experiments with Quantinuum's \texttt{H1-2} Hardware}
Our second set of results considers linear systems of the form in section \ref{sec:simplest} that can be reformulated as Forrelation problems.
Consequently, the quantum-executed Woodbury approach is provably faster than classical approaches.
We tested this on a suite of problems that have Forrelation that is either positive or negative and is not close to zero (so that a noisy computation will not get the right answer for the wrong reasons).
These circuits were generated using a search algorithm to find circuits with Forrelation easily distinguishable from zero.

We used Quantinuum's \texttt{H1-2} machine with 12 qubits, and---as in section \ref{sec:results_ibm}---utilized the Hadamard test and not the swap test.
The \texttt{H1-2} device is a trapped-ion based quantum computer with a fully connected hardware graph, meaning that no SWAP gates need to be used when compiling the circuits to the hardware\cite{Pino_2021}. 
Each inner product was estimated using $10^3$ shots; again, for convenience, we did not utilize an adaptive sampling procedure when determining the number of shots. 
The largest system size that can be solved on the \texttt{H1-2} device is $2^{11}$ equations/unknowns, which corresponds to a circuit with $12$ qubits, which is the maximum available number of qubits available on the device at the time these experiments were run.\footnote{These quantum circuits were run on Quantinuum \texttt{H1-2} in September 2022.}
The circuits were submitted to \texttt{H1-2} as an OpenQASM 2.0 string \cite{cross2017open} that was compiled for an arbitrary connectivity (e.g., assuming a fully connected hardware graph), and server-side compilation of the circuits adpated the circuits to the hardware-compatible gateset.
To contextualize the size of the circuits, we note that for linear systems of size $N=2^{11}$ contains $99$ two qubit gates.

\begin{figure}[ht]
\begin{centering}
    \includegraphics[width=0.7\textwidth]{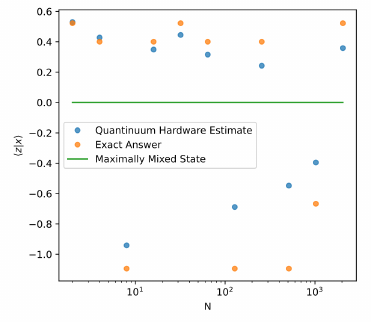}
    \caption{Estimates of $\braket{z}{x}$ for problems ranging in size from two equations/unknowns to $2^{11}$ equations/unknowns using our algorithm on Quantinuum's \texttt{H1-2} hardware. The blue dots illustrate the Quantinuum \texttt{H1-2} estimates of the orange dots, which are the true solutions. The green line is the maximally-mixed state, or a state in which any `information' from the quantum computer has been reduced to random noise.
    }
    \label{fig:quantinuum_results}
\end{centering}
\end{figure}

Figure \ref{fig:quantinuum_results} illustrates the results for linear systems of size $N$ between two and $2^{11}$, in which the estimated inner products were plugged into equation \ref{eq:case1final} to obtain the set of points labelled ``Quantinuum Hardware Estimate.''\footnote{The maximum $N$ used in this case is smaller than in section \ref{sec:results_ibm} solely as a consequence of the available number of qubits when these experiments were run.}
Again, the results are promising, with all equations solved to a relative error within 50\%, and the first half of systems solved to an error within 21.5\%.
While these errors are larger than those for the comparably-sized problems of section \ref{sec:results_ibm}, the fact that we obtain such accurate solutions with no error mitigation and with $10^2$ fewer shots than in section \ref{sec:results_ibm} is notable.
This is particularly true given the distance of the true solutions from the maximally-mixed state, which is a quantum state reduced entirely to random noise \cite{nielsen2010quantum}.
Notice that all of the true solutions are far from the maximally-mixed state; this suggests that the information we extract from the quantum computer is a genuine reflection of the hardware's ability to accurately execute the Woodbury-approach circuits, rather than effects of noise that happen to align with desired results from the prepared solutions.

\section{Conclusion}
We have described a quantum algorithm for solving linear systems of equations that fit within the paradigm of the Woodbury identity.
That is, the matrix representing the linear system of equations must be a low-rank modification of a system of equations that can be solved efficiently, such as a unitary matrix.
The low-rank aspect is key since it keeps the amount of information that must be extracted from the quantum computer small.
While limited to such systems, this approach has advantages over existing quantum linear systems algorithms.
Namely, it only relies on the Hadamard test, which makes it much simpler and more applicable in the near term than HHL and improvements thereof.
This dramatically improves the constant factors that are discarded in a scaling analysis but are important in applying the algorithm, especially on noisy hardware.
While the worst-case scaling in the condition number is worse for this approach than for HHL, for many practical applications the scaling is equivalent or significantly better.
It also does not require a variational optimization loop, which reduces the cost compared to VQLS and avoids problems with local minima or barren plateaus.
Thinking long term, combining this approach with other algorithms such as HHL or VQLS (for solving the $A^{-1}$ part of equation \ref{eq:woodbury}) could extend the reach of those algorithms to include the low-rank modifications. 
More research is required to explore any potential benefits of combining the Woodbury approach with other quantum algorithms for linear systems.

A key question is whether or not this can be implemented more efficiently on a quantum computer than on a classical computer.
This question has previously been considered in the context where the vectors are stored in quantum RAM \cite{aaronson2015read,lloyd2013quantum}, which differs somewhat from the context here where unitaries prepare the vectors.
Answering this question requires understanding how long it takes to compute inner products on classical and quantum computers. Hence, the answer requires more information about the problem being solved.
If the individual components of each of the vectors in the inner product can be computed in $O(\log N)$ on a classical computer, the inner products can be estimated by sampling.
In cases such as this, it is unlikely that a quantum computer could outperform a classical computer.
However, there may be cases where it is expensive to compute the components individually, and they are more naturally computed all at once (e.g., in training a machine learning model).
In these cases, the classical computer requires at least $O(N)$ operations to compute the components.
Another more contrived case where quantum advantage might exist would be when the unitaries, $U_i$, are designed to be hard to simulate classically (e.g., quantum supremacy circuits).
So, there is potential for the inner products to be computed more efficiently on a quantum computer, where the circuit depth of the unitaries scale like $O(\log N)$.

Our experimental investigation on quantum computing hardware indicates the potential for good performance in the near term.
The largest system we have seen solved on a quantum computer previously involves $2^{17}$ equations and was solved with much lower accuracy \cite{perelshtein2020large}.
So, these results push the frontier of size/accuracy for quantum linear systems algorithms.
It is noteworthy that zero noise (linear) extrapolation significantly improved the algorithm's performance, reducing the error by more than a factor of $10$.
Alternative error mitigation techniques or more advanced zero noise extrapolation could potentially be used to improve the performance further.
Studying the Woodbury algorithm with these alternative error mitigation techniques presents an interesting avenue for further work.

\section*{Data and code availability}
The code and data used to generate the figures is available at \url{https://github.com/omalled/quantum-woodbury}.

\section*{Acknowledgements}
This work was supported by the U.S. Department of Energy through the Los Alamos National Laboratory.
Los Alamos National Laboratory is operated by Triad National Security, LLC, for the National Nuclear Security Administration (NNSA) of U.S. Department of Energy (Contract No. 89233218CNA000001).
DO, JMH, and JG acknowledge support from Los Alamos National Laboratory's Laboratory Directed Research and Development program through project 20220077ER and 
YS acknowledges support from Los Alamos National Laboratory's Laboratory Directed Research and Development program through project 20210639ECR.
EP, RL, and SE acknowledge support from the NNSA's Advanced Simulation and Computing Beyond Moore's Law Program at Los Alamos National Laboratory.
SG acknowledges support from the United States Department of Energy through the Computational Science Graduate Fellowship (DOE CSGF) under grant number DE-SC0019323.
This research used resources of the Oak Ridge Leadership Computing Facility, which is a DOE Office of Science User Facility supported under Contract DE-AC05-00OR22725. The Oak Ridge Leadership Computing Facility provided access to the Quantinuum H1-2 computer. 
This research also relied on the use of IBM Quantum services through the IBMQ-LANL Quantum Hub. 

\appendix

\section{Relative error analysis}\label{app:error-analysis}
Here, we provide an analysis of the number of shots to estimate $\braket{z}{\tilde{x}}$ from section \ref{sec:simplest} with relative error $\epsilon$, where
\begin{equation}
   \ket{\tilde{x}} = \frac{\ket{x}}{\sqrt{\braket{x}{x}}} 
\end{equation}
is the normalized version of $\ket{x}$.
This allows for a more straightforward comparison of the Woodbury approach with other quantum algorithms, which typically compute the normalized solution.
For example, the original HHL algorithm prepares a normalized state with error $\epsilon$ in a runtime of $O(\kappa^2 \log(N)/\epsilon)$, where $\kappa$ is the condition number of the matrix $A$.
We are particularly interested in how the number of shots scales with $\kappa$ for the Woodbury approach.
As discussed previously, the large $\kappa$ limit generally represents the most difficult to solve linear systems, thus providing a useful worst-case-scenario study of algorithmic performance.

In this section we consider the rank-1 case of $A = I + UV$ where $U = \alpha_0\ket{u_0}$ and $V = \beta_0\bra{v_0}$.
We will assume that all vectors ($\ket{z}$, $\ket{b}$, $\ket{v_0}$, $\ket{u_0}$) are real and $\alpha_0$, $\beta_0 > 0$.
To simplify the notation, we will drop the subscripts ($\ket{v_0}\rightarrow\ket{v}$ and $\ket{u_0}\rightarrow\ket{u}$) and denote $\alpha\equiv \alpha_0\beta_0$.
With this notation we have $A = I + \alpha \ket{u}\bra{v}$.
Applying corrolary \ref{coro:kappa} from appendix \ref{app:cond_number} and writing $\cos\theta$ as $\braket{v}{u}$, the condition number of $A$ is given by
\begin{equation}\label{eq:kappa}
    \kappa = \frac{\sqrt{\alpha^2 / 2 + \alpha\braket{v}{u} + 1 + \alpha/2 \sqrt{\alpha^2+4\alpha\braket{v}{u} + 4}}}{\sqrt{\alpha^2 / 2 + \alpha\braket{v}{u} + 1 - \alpha/2 \sqrt{\alpha^2+4\alpha\braket{v}{u} + 4}}}.
\end{equation}
Straightforward but tedious algebraic manipulation shows that solving equation \ref{eq:kappa} to represent $\braket{v}{u}$ in terms of $\kappa$ and $\alpha$ gives two solutions, which we label $\braket{v}{u}^+$ and $\braket{v}{u}^-$:
\begin{equation}\label{eq:vu}
    \braket{v}{u}^{\pm} = \pm \frac{\kappa(\alpha^2 \mp \kappa +2)\mp 1}{\alpha(\kappa \mp 1)^2}.
\end{equation}
The two solutions arise because inverting equation \ref{eq:kappa} requires solving a quadratic expression in $\braket{v}{u}$.
Note that because $-1\le\braket{v}{u}\le 1$, equation \ref{eq:kappa} has several additional restrictions depending on the value of $\alpha$:
\begin{equation}
    \begin{tabular}{c|c|c}    
        & $\braket{v}{u}^-$ validity & $\braket{v}{u}^+$ validity \\
       \hline
       $0<\alpha<1$  & none  & $1+\alpha\leq \kappa \leq \frac{1}{1-\alpha}$  \\
       \hline
       $\alpha=1$ & none & $\kappa \ge \alpha + 1$ \\
       \hline
       $1<\alpha<2$ & $\kappa \geq \frac{1}{\alpha-1}$ & $\kappa \ge \alpha + 1$ \\
       \hline
       $\alpha\ge 2$ & $\kappa \ge \alpha-1$ & $\kappa \ge \alpha + 1$
    \end{tabular} 
\end{equation}
We can also see from equation \ref{eq:vu} that in the limit $\kappa \to \infty$, $\braket{v}{u} \to -\frac{1}{\alpha}$; $\braket{v}{u}^+$ approaches from above and $\braket{v}{u}^-$ approaches from below.

Let us also study equation \ref{eq:kappa} in the Hermitian case, i.e., when $\braket{v}{u} = \pm 1$:
\begin{equation}
    \kappa = \begin{cases}
        \alpha + 1, & \braket{v}{u} = +1,\\
            \frac{1}{|\alpha - 1|}, & \braket{v}{u} = -1 \text{ and } 0<\alpha <2,\\
            \alpha - 1, & \braket{v}{u} = -1 \text{ and } \alpha \ge 2.
    \end{cases}
\end{equation}

Having established how the quantities of $\braket{v}{u}$, $\alpha$, and $\kappa$, are related, we can turn to our analysis of the scaling performance of the relative error in the Woodbury approach.
Recall that the unnormalized solution to $Ax = b$ is given by
\begin{equation}
    \ket{x} = \ket{b} - \frac{\alpha\braket{v}{b}}{1+\alpha\braket{v}{u}}\ket{u}.
\end{equation}
Normalizing gives us
\begin{eqnarray}
    \braket{x}{x}
    &=& \braket{b}{b} - \frac{2\alpha\braket{v}{b}\braket{b}{u}}{1+\alpha\braket{v}{u}} + \left(\frac{\alpha\braket{v}{b}}{1+\alpha\braket{v}{u}}\right)^2\braket{u}{u} \\
    &=& 1 - \frac{2\alpha\braket{v}{b}\braket{b}{u}}{1+\alpha\braket{v}{u}} + \left(\frac{\alpha\braket{v}{b}}{1+\alpha\braket{v}{u}}\right)^2
\end{eqnarray}
so,
\begin{eqnarray}
    \ket{\tilde{x}} 
    &=&
    \frac{\ket{b} - \frac{\alpha\braket{v}{b}}{1 + \alpha\braket{v}{u}}\ket{u}}{\sqrt{1 - \frac{2\alpha\braket{v}{b}\braket{b}{u}}{1+\alpha\braket{v}{u}} + \left(\frac{\alpha\braket{v}{b}}{1+\alpha\braket{v}{u}}\right)^2}}
\end{eqnarray}
and,
\begin{eqnarray}\label{eq:zx_norm}
    \braket{z}{\tilde{x}} 
    &=& 
    \frac{\braket{z}{b} - \frac{\alpha\braket{v}{b}}{1 + \alpha\braket{v}{u}}\braket{z}{u}}{\sqrt{1 - \frac{2\alpha\braket{v}{b}\braket{b}{u}}{1+\alpha\braket{v}{u}} + \left(\frac{\alpha\braket{v}{b}}{1+\alpha\braket{v}{u}}\right)^2}}.
\end{eqnarray}
Note that in addition to estimating $\braket{z}{b}$, $\braket{v}{b}$, $\braket{v}{u}$, and $\braket{z}{u}$ (which are needed for estimating $\braket{z}{x}$), we must also estimate the real part of $\braket{b}{u}$ in order to estimate $\braket{z}{\tilde{x}}$.
As in section \ref{sec:simplest}, we can determine the number of shots necessary to estimate equation \ref{eq:zx_norm} with relative error $\epsilon$.
To do so, note that the Hadamard test requires $O(\frac{1}{\epsilon^2})$ shots to obtain a result with a relative error of $\epsilon$.
So, obtaining the entirety of equation \ref{eq:zx_norm} with a relative error of $\epsilon$ requires weighting the number of shots required to obtain each constituent inner product with a coefficient that reflects that inner product's effect on equation \ref{eq:zx_norm}.
To obtain those coefficients, apply a Taylor expansion -- centered about approximate values of each of the inner products -- to the right-hand side of equation \ref{eq:zx_norm}, which gives
\begin{equation}
\begin{split}
    \braket{z}{\tilde{x}} 
    =
    \frac{\partial{\braket{z}{\tilde{x}}}}{\partial{\braket{z}{b}}}\left(\braket{z}{b} - \braket{z}{b}_{approx}\right) +
    \frac{\partial{\braket{z}{\tilde{x}}}}{\partial{\braket{v}{b}}}\left(\braket{v}{b} - \braket{v}{b}_{approx}\right) +
    \frac{\partial{\braket{z}{\tilde{x}}}}{\partial{\braket{v}{u}}}\left(\braket{v}{u} - \braket{v}{u}_{approx}\right) +
    \\
    \frac{\partial{\braket{z}{\tilde{x}}}}{\partial{\braket{z}{u}}}\left(\braket{z}{u} - \braket{z}{u}_{approx}\right) +
    \frac{\partial{\braket{z}{\tilde{x}}}}{\partial{\braket{b}{u}}}\left(\braket{b}{u} - \braket{b}{u}_{approx}\right) + O\left(...^2\right)
    .
\end{split}
\end{equation}
The coefficients (i.e., partial derivatives) for each term are multiplied by $\frac{1}{\epsilon}$ and are squared according to the shot count scaling for the Hadamard test.
Thus, computing the partial derivatives gives the following number of shots required to compute each inner product such that the value of equation \ref{eq:zx_norm} has relative error $\epsilon$:
\begin{equation}\label{eq:err-zb}
    N_{\braket{z}{b}} = \frac{1}{\epsilon ^2 \left(\frac{\alpha  \braket{v}{b} (\alpha  \braket{v}{b}-2 \braket{b}{u} (\alpha  \braket{v}{u}+1))}{(\alpha  \braket{v}{u}+1)^2}+1\right)}\,.
\end{equation}
\begin{widerequation}\label{eq:err-vb}
    N_{\braket{v}{b}} =  \frac{\alpha ^4 (1+\alpha  \braket{v}{u})^2 (\braket{v}{b} \braket{z}{b}+\braket{v}{u} \braket{z}{u}+\braket{z}{u}/\alpha-\braket{b}{u} (\braket{v}{b} \braket{z}{u}+\braket{v}{u} \braket{z}{b}+\braket{z}{b}/\alpha))^2}{\epsilon ^2 \left(\alpha  \left(-2 \braket{b}{u} (\alpha  \braket{v}{b} \braket{v}{u}+\braket{v}{b})+\alpha  \left(\braket{v}{b}^2+\braket{v}{u}^2\right)+2 \braket{v}{u}\right)+1\right)^3}\,.
\end{widerequation}
\begin{widerequation}\label{eq:err-vu}   
    N_{\braket{v}{u}} = \frac{\alpha ^4 \braket{v}{b}^2 (-\braket{b}{u} (\alpha  \braket{v}{b} \braket{z}{u}+\alpha  \braket{v}{u} \braket{z}{b}+\braket{z}{b})+\alpha  \braket{v}{b} \braket{z}{b}+\alpha  \braket{v}{u} \braket{z}{u}+\braket{z}{u})^2}{\epsilon ^2 \left(\alpha  \left(-2 \braket{b}{u} (\alpha  \braket{v}{b} \braket{v}{u}+\braket{v}{b})+\alpha  \left(\braket{v}{b}^2+\braket{v}{u}^2\right)+2 \braket{v}{u}\right)+1\right)^3}\,. 
\end{widerequation}
\begin{equation}\label{eq:err-zu}
    N_{\braket{z}{u}} = \frac{\alpha ^2 \braket{v}{b}^2}{\epsilon ^2 \left(\alpha  \left(-2 \braket{b}{u} (\alpha  \braket{v}{b} \braket{v}{u}+\braket{v}{b})+\alpha  \left(\braket{v}{b}^2+\braket{v}{u}^2\right)+2 \braket{v}{u}\right)+1\right)}\,. 
\end{equation}
\begin{equation}\label{eq:err-bu}
    N_{\braket{b}{u}} = \frac{\alpha ^2 \braket{v}{b}^2 (\alpha  \braket{v}{u}+1)^2 (-\alpha  \braket{v}{b} \braket{z}{u}+\alpha  \braket{v}{u} \braket{z}{b}+\braket{z}{b})^2}{\epsilon ^2 \left(\alpha  \left(-2 \braket{b}{u} (\alpha  \braket{v}{b} \braket{v}{u}+\braket{v}{b})+\alpha  \left(\braket{v}{b}^2+\braket{v}{u}^2\right)+2 \braket{v}{u}\right)+1\right)^3}\, . 
\end{equation}

Before continuing our analysis, we make the following noteworthy simplifying assumption: we assume that $z$ and $b$ are neither exactly parallel or orthogonal to $v$ or $u$, meaning the inner products $\braket{v}{b}, \braket{z}{u}, \braket{b}{u}$ are not exactly $0$ or $\pm 1$.
(They can be arbitrarily close to parallel/orthogonal, just not exactly so.)
When these assumptions are not met, intricate cancellations can occur in equations ~\ref{eq:err-zb}-\ref{eq:err-bu}, which can produce significantly worse scaling in $\kappa$.
We do not wish to minimize or obfuscate the existence of these niche scenarios, and we will highlight examples below.
However the focus of our analysis will assume we are free of such specific state overlaps.

Let us now work through some special cases of the above equations, beginning with the Hermitian case.
As discussed previously there are really three distinct subcases here: $\braket{v}{u} = 1$, $\braket{v}{u} = -1$ with $0<\alpha<2$, and $\braket{v}{u} = -1$ with $\alpha>2$.
These cases assume that we know that $A = I + \alpha \ket{u}\bra{v}$ is Hermitian and the value of $\alpha$ \emph{a priori}. 
Therefore, determining which of the above cases is relevant requires only a single Hadamard test on $\braket{v}{u}$.
For the first case, $\braket{v}{u} = 1$, we have $\alpha = \kappa-1$ and equations~\ref{eq:err-zb}-\ref{eq:err-bu} in the $\kappa\to\infty$ limit simplify to
\begin{eqnarray}
    \lim_{\kappa \to\infty}N_{\braket{z}{b}} &=&  \frac{1}{\epsilon ^2(1-\braket{v}{b}^2)}, \\
    \lim_{\kappa \to\infty}N_{\braket{v}{b}} &=&  \frac{\braket{z}{u}^2}{\epsilon ^2\left(1-\braket{v}{b}^2\right)}, \\
    \lim_{\kappa \to\infty}N_{\braket{v}{u}} &=&  1, \\
    \lim_{\kappa \to\infty}N_{\braket{z}{u}} &=&  \frac{\braket{v}{b}^2}{ \epsilon ^2\left(1-\braket{v}{b}^2\right)}, \\
    \lim_{\kappa \to\infty}N_{\braket{b}{u}} &=&  \frac{\braket{v}{b}^2 (\braket{z}{b}-\braket{v}{b} \braket{z}{u})^2}{\epsilon ^2\left(1-\braket{v}{b}^2\right)^3 }. 
\end{eqnarray}
Critically, with our assumption that $\braket{v}{b} \ne \pm1$,  none of these terms diverge as $\kappa\to\infty$, a sharp contrast from HHL, which diverges as $O(\kappa^2/\epsilon)$.
A similar story emerges for the $\braket{v}{u} = -1$ case: the number of shots required for $\epsilon$ error asymptotes to a constant as $\kappa$ diverges.
However, it is worth emphasizing that when our assumption about $\braket{v}{b}$ is not true, meaning that $\braket{v}{b} = \pm 1$, the Woodbury approach instead scales as $O(\kappa^4/\epsilon^2)$.

We now turn to the non-Hermitian case with $\alpha$ still equal to $1$.
In this case equations~\ref{eq:err-zb}-\ref{eq:err-bu} in the $\kappa\to\infty$ limit simplify to
\begin{eqnarray}
    \lim_{\kappa \to\infty} N_{\braket{z}{b}} &=& \lim_{\kappa \to\infty} N_{\braket{v}{b}} = \lim_{\kappa \to\infty} N_{\braket{b}{u}} = 0. \label{eq:nvb_zero}\\
    \lim_{\kappa \to\infty} N_{\braket{z}{u}} &=&  \frac{1}{\epsilon ^2}. \\
    \lim_{\kappa \to\infty} N_{\braket{v}{u}} &=&  \frac{(\braket{z}{b}-\braket{b}{u} \braket{z}{u})^2}{\epsilon ^2\braket{v}{b}^2}.
\end{eqnarray}
Again we see that the number of shots for the Woodbury approach asymptotes to a constant as $\kappa\to\infty$ (the case where $\braket{v}{b} = 0$ appears troublesome but in fact has finite scaling as well). 
This is because in this limit, $\braket{z}{\tilde{x}}\to\text{sgn}(\braket{v}{b})\braket{z}{u}$.
The fact that $\lim_{\kappa \to\infty} N_{\braket{v}{b}} = 0$ in equation \ref{eq:nvb_zero} is misleading, as a fixed number of shots (i.e., independent of $\epsilon$) is required to accurately determine $\text{sgn}(\braket{v}{b})$. 

Finally, consider the general case of a non-Hermitian matrix with $\alpha>1$.
Here we are treating $\alpha$ as a fixed and known quantity and only considering the large $\kappa$ limit.
In the previous cases studied, we could simply plug-in an expression for $\braket{v}{u}$ in terms of $\kappa$ and $\alpha$ into equations~\ref{eq:err-zb}-\ref{eq:err-bu} and take the $\kappa\to\infty$ limit. 
This was possible because in each case there was only one valid substitution for $\braket{v}{u}$.
However, in the general case we have both $\braket{v}{u}^+$ and $\braket{v}{u}^-$.
Knowing which of the $\braket{v}{u}^\pm$ solutions to subsitute is equivalent to knowing the value of sgn$(1+\alpha \braket{v}{u})$, since $\braket{v}{u}^+ \ge -1/\alpha$ and $\braket{v}{u}^- \le -1/\alpha$, and this is exactly the quantity we need to determine experimentally.
This is clear when looking at the very large (but finite) $\kappa$ regime of equation \ref{eq:zx_norm},
\begin{equation}
    \braket{z}{\tilde{x}}\approx \text{sgn}(1+\alpha \braket{v}{u})\text{sgn}(\braket{v}{b})\braket{z}{u}.
\end{equation}
(In the strict $\kappa=\infty$ limit, $\braket{z}{\tilde{x}}$ is ill-defined.)
Putting either $\braket{v}{u}^+$ or $\braket{v}{u}^-$ into equations~\ref{eq:err-zb}-\ref{eq:err-bu} and taking the $\kappa\to\infty$ limit gives a finite number of shots, which is clearly a nonsensical result, as accurately determining $\text{sgn}(1+\alpha \braket{v}{u})$ as $\braket{v}{u}\to -1/\alpha$ will necessarily require an increasingly large number of shots.
In particular, for large $\kappa$,
\begin{equation}
    \braket{v}{u}^{\pm}\approx -\frac{1}{\alpha} \pm \frac{\alpha}{\kappa}
\end{equation}
Therefore the accuracy with which we must determine $\braket{v}{u}$, in order to accurately determine $\text{sgn}(1+\alpha \braket{v}{u})$, increases as $O(\kappa)$.
Since the Hadamard test scales as $O(1/\epsilon^2)$, we conclude that the relative error of the Woodbury approach in the general case scales as $O(\kappa^2/\epsilon^2)$.
However, in certain cases (e.g., Hermitian matrices and matrices with $\alpha=1$ and simplifying assumptions on $\braket{v}{b}, \braket{z}{u}, and \braket{b}{u}$), the Woodbury approach with rank-1 matrices has a performance scaling that does not diverge with $\kappa$.
This all said, it is worth again emphasizing our simplifying assumptions that none of $\braket{v}{b}, \braket{z}{u}, \braket{b}{u}$ are exactly $0$ or $\pm 1$. 
If these assumptions are not met then more nuanced analysis is required, and the Woodbury approach can scale as $O(\kappa^4/\epsilon^2)$. 

\section{Analysis of the condition number}
\label{app:cond_number}
\begin{lemma}
If $\mathbf{a}, \mathbf{b} \in \mathbb{R}^2$ then the singular values of $I+\mathbf{a}\mathbf{b}^T$ are 
\begin{eqnarray}
    \sigma_1 &=& \sqrt{x + y / 2} \label{eq:sigma1} \\
    \sigma_2 &=& \sqrt{x - y / 2} \label{eq:sigma2} \\
    x &=&  ||\mathbf{a}||^2||\mathbf{b}||^2 / 2 + \mathbf{a}\cdot\mathbf{b} + 1 \label{eq:x}\\
    y &=& ||\mathbf{a}||||\mathbf{b}|| \sqrt{||\mathbf{a}||^2||\mathbf{b}||^2+4\mathbf{a}\cdot\mathbf{b} + 4} \label{eq:y}
\end{eqnarray}
\end{lemma}

\begin{proof}
Let $A$ be a $2\times 2$ matrix,
\begin{equation}
    A = \begin{pmatrix} a & b\\
    c & d
    \end{pmatrix}.
\end{equation}
The singular values are the square roots of the eigenvalues of $A^TA$.
\begin{equation}
    A^TA = \begin{pmatrix} 
     a^2+c^2 & a b+c d \\
     a b+c d & b^2+d^2 \\
    \end{pmatrix}
\end{equation}
Setting $|A^TA-\lambda I|=0$ gives 
\begin{equation}
    \lambda ^2 -\lambda  \left(a^2+b^2+c^2+d^2\right)+(a d-b c)^2 = 0,
\end{equation}
which has solutions
\begin{equation}
    \lambda = \frac{S_1 \pm S_2}{2},
\end{equation}
where
\begin{equation}\label{eq:s_vals}
    S_1 = a^2 + b^2 + c^2 + d^2, S_2 = \sqrt{S_1^2-4(ad-bc)^2}.
\end{equation}
Therefore the singular values of $A$ are 
\begin{equation}
    \sigma_1 = \sqrt{\frac{S_1+S_2}{2}}, \sigma_2 = \sqrt{\frac{S_1-S_2}{2}}.
    \label{eq:sigmas}
\end{equation}
In the case of $A = I+\mathbf{a}\mathbf{b}^T$ for $\mathbf{a}, \mathbf{b} \in \mathbb{R}^2$, we have $\mathbf{a} = (a_1~a_2)$, $\mathbf{b} = (b_1~b_2)$, and 
\begin{equation}
    A = \begin{pmatrix} 
 1+a_1 b_1 & a_1 b_2 \\
 a_2 b_1 & 1+a_2 b_2 \\
    \end{pmatrix}.
\end{equation}
Inserting these values into equation \ref{eq:s_vals} gives
\begin{eqnarray}
    S_1 = \|\mathbf{a}\|^2\|\mathbf{b}\|^2+2\mathbf{a}\cdot\mathbf{b}+2,\\
    S_2 = \|\mathbf{a}\|\|\mathbf{b}\|\sqrt{\|\mathbf{a}\|^2\|\mathbf{b}\|^2 + 4\mathbf{a}\cdot\mathbf{b} + 4 },
\end{eqnarray}
Note that $x=S_1/2$ and $y=S_2$ in equations \ref{eq:x} and \ref{eq:y}, respectively.
Inserting this into equation \ref{eq:sigmas} gives the desired result.
\end{proof}

\begin{theorem}
If $\mathbf{a}, \mathbf{b} \in \mathbb{R}^N$ then the singular values of $I+\mathbf{a}\mathbf{b}^T$ are all one with the possible exception of two singular values, which are as given in equations \ref{eq:sigma1}--\ref{eq:y}.
\label{thm:rank1_svd}
\end{theorem}
\begin{proof}
Let $\mathbf{u}_i$ for $i=1,\ldots,N$ be an orthonormal basis for $\mathbb{R}^N$ such that $\mathbf{a}$ and $\mathbf{b}$ are in the subspace spanned by $\mathbf{u}_1$ and $\mathbf{u}_2$. Let $\mathbf{a}=a_1\mathbf{u}_1+a_2\mathbf{u}_2$ and $\mathbf{b}=b_1\mathbf{u}_1+b_2\mathbf{u}_2$. Then
\begin{eqnarray}
    I+\mathbf{a}\mathbf{b}^T
    &=& \sum_{i=1}^N \mathbf{u}_i\mathbf{u}_i^T + (a_1\mathbf{u}_1+a_2\mathbf{u}_2) (b_1\mathbf{u}_1+b_2\mathbf{u}_2)^T  \\
    &=& \sum_{i=3}^N \mathbf{u}_i\mathbf{u}_i^T + (1 + a_1 b_1)\mathbf{u}_1\mathbf{u_1}^T + (1 + a_2 b_2)\mathbf{u}_2\mathbf{u_2}^T  \\
    && + a_1 b_2 \mathbf{u}_1 \mathbf{u}_2^T + a_2 b_1 \mathbf{u}_2 \mathbf{u}_1^T \\
    &\equiv& \sum_{i=3}^N \mathbf{u}_i\mathbf{u}_i^T + M \label{eq:uuM}
\end{eqnarray}

We now concern ourselves with $M=(1 + a_1 b_1)\mathbf{u}_1\mathbf{u_1}^T + (1 + a_2 b_2)\mathbf{u}_2\mathbf{u_2}^T + a_1 b_2 \mathbf{u}_1 \mathbf{u}_2^T + a_2 b_1 \mathbf{u}_2 \mathbf{u}_1^T$, which is close to the form needed to apply Lemma 1, except that it is embedded in $\mathbb{R}^N$ rather than $\mathbb{R}^2$. Let $\mathbf{e}_1=[1, 0]$ and $\mathbf{e}_2=[0, 1]$ be the two standard basis vectors in $\mathbb{R}^2$. We multiply $M$ on the left by $L=\mathbf{e}_1\mathbf{u}_1^T+\mathbf{e}_2\mathbf{u}_2^T$ and on the right by $L^T$. Note that $L$ maps between the subspace spanned by ${\mathbf{u}_1,\mathbf{u}_2}$ and $\mathbb{R}^2$, which enables us to apply Lemma 1
\begin{eqnarray}
    LML^T
    &=& I+[a_1, a_2]^T[b_1, b_2] \\
    &=& U \Sigma V^T \label{eq:svd}
\end{eqnarray}
where $\Sigma$ is a diagonal matrix with diagonal entries $\sigma_1$ and $\sigma_2$ from Lemma 1, $U$ and $V$ are unitary matrices, and $V^T$ denotes the conjugate transpose of $V$.

Observe that
\begin{equation}
    L^TLML^TL=M \label{eq:llmll}
\end{equation}
which can be verified by checking three cases: 1) applying either matrix to $\mathbf{u}_i$ for $i\ge 3$ is zero, 2) applying either matrix to $\mathbf{u}_1$ is $(1+a_1 b_1)\mathbf{u}_1+a_2 b_1 \mathbf{u}_2$, and 3) applying either matrix to $\mathbf{u}_2$ is $(1+a_2 b_2)\mathbf{u}_2+a_1 b_2 \mathbf{u}_1$. Since both are linear and operate in the same fashion on the $\mathbf{u}_i$ basis, they are equal.

Combining equations \ref{eq:uuM}, \ref{eq:svd}, and \ref{eq:llmll} gives
\begin{equation}
    I + \mathbf{a}\mathbf{b}^T = \sum_{i=3}^N \mathbf{u}_i\mathbf{u}_i^T + L^TU\Sigma V^TL \label{eq:almost_there}
\end{equation}
Let $\tilde{\mathbf{u}}_1$ and $\tilde{\mathbf{u}}_2$ be the first and second columns of $L^TU$ and $\tilde{\mathbf{v}}_1$ and $\tilde{\mathbf{v}}_2$ be the first and second columns of $L^TV$. Substituting this into equation \ref{eq:almost_there} gives
\begin{equation}
    I + \mathbf{a}\mathbf{b}^T = \sum_{i=3}^N \mathbf{u}_i\mathbf{u}_i^T + \sum_{i=1}^2 \tilde{\mathbf{u}}_i \sigma_i \tilde{\mathbf{v}}_i^T
\end{equation}
This is in the form of a singular value decomposition with the desired singular values.

It remains to show that $\{\tilde{\mathbf{u}}_1, \tilde{\mathbf{u}}_2, \mathbf{u}_3, \ldots, \mathbf{u}_N \}$ and $\{\tilde{\mathbf{v}}_1, \tilde{\mathbf{v}}_2, \mathbf{v}_3, \ldots, \mathbf{v}_N \}$ are the columns of unitary matrices. By construction of the basis, $\mathbf{u}_i^T\mathbf{u}_j=\delta_{ij}$. Also, $\tilde{\mathbf{u}}_i^T\tilde{\mathbf{u}}_j=\delta_{ij}$ since $(L^TU)^T(L^TU)=U^TLL^TU=U^TU=I$.
Finally, $\tilde{\mathbf{u}}_i\mathbf{u}_j=0$ for $j \ge 3$ since $L\mathbf{u}_j=0$ when $j\ge 3$. This shows that $\{\tilde{\mathbf{u}}_1, \tilde{\mathbf{u}}_2, \mathbf{u}_3, \ldots, \mathbf{u}_N \}$ are the columns of a unitary matrix. Essentially the same argument shows that $\{\tilde{\mathbf{v}}_1, \tilde{\mathbf{v}}_2, \mathbf{v}_3, \ldots, \mathbf{v}_N \}$ are the columns of a unitary matrix.
\end{proof}

\begin{corollary}\label{coro:kappa}
    If $\mathbf{a} = \alpha_0\mathbf{a_0},\mathbf{b}=\beta_0\mathbf{b_0}\in\mathbb{R}^N$, and $||\mathbf{a_0}||=||\mathbf{b_0}||=1$, the condition number, $\kappa$, of $I+\mathbf{a}\mathbf{b}^T$ is
    \begin{equation}
        \kappa = \frac{\sqrt{\alpha^2 / 2 + \alpha\cos\theta + 1 + \alpha/2 \sqrt{\alpha^2+4\alpha\cos\theta + 4}}}{\sqrt{\alpha^2 / 2 + \alpha\cos\theta + 1 - \alpha/2 \sqrt{\alpha^2+4\alpha\\cos\theta + 4}}},
    \end{equation}
    where $\theta$ is the angle between $\mathbf{a_0}$ and $\mathbf{b_0}$, and $\alpha=\alpha_0\beta_0$.
    Additionally, if $\mathbf{a}$ and $\mathbf{b}$ are normalized, meaning $\alpha=1$, then
    \begin{equation}
        \kappa = \frac{\sqrt{3+2\cos\theta + \sqrt{5+4\cos\theta}}}{\sqrt{3+2\cos\theta - \sqrt{5+4\cos\theta}}}.
    \end{equation}
\end{corollary}
\begin{proof}
Substituting  $\mathbf{a}\cdot\mathbf{b}= \alpha\cos\theta$ and $||\mathbf{a}||||\mathbf{b}||=\alpha$ into equations \ref{eq:sigma1}--\ref{eq:y} and simplifying gives
\begin{equation}\label{eq:general_kappa_def}
\frac{\sigma_1}{\sigma_2} = \frac{\sqrt{\alpha^2 / 2 + \alpha\cos\theta + 1 + \alpha/2 \sqrt{\alpha^2+4\alpha\cos\theta + 4}}}{\sqrt{\alpha^2 / 2 + \alpha\cos\theta + 1 - \alpha/2 \sqrt{\alpha^2+4\alpha\\cos\theta + 4}}} .
\end{equation}
By theorem \ref{thm:rank1_svd}, the set of singular values of $I+\mathbf{a}\mathbf{b}^T$ is $\{1, \sigma_1, \sigma_2\}$ when $N>2$ and $\{\sigma_1, \sigma_2\}$ when $N=2$. Since the condition number is the ratio of the largest to the smallest singular value, it remains to show that $\sigma_1\ge1$ and $\sigma_2\le 1$.
We begin by showing that $\sigma_1^2\geq 1$:
\begin{eqnarray*}
\sigma_1^2 &=& x + y/2\\
&=& ||\mathbf{a}||^2||\mathbf{b}||^2/2 + \mathbf{a}\cdot\mathbf{b} + 1 + \frac{1}{2}||\mathbf{a}||||\mathbf{b}|| \sqrt{||\mathbf{a}||^2||\mathbf{b}||^2 + 4\mathbf{a}\cdot\mathbf{b} + 4}\\
&\geq& (\mathbf{a}\cdot\mathbf{b})^2/2 + \mathbf{a}\cdot\mathbf{b} + 1 + \frac{1}{2}|\mathbf{a}\cdot \mathbf{b}| \sqrt{(\mathbf{a}\cdot\mathbf{b})^2 + 4\mathbf{a}\cdot\mathbf{b} + 4}\\
&=& (\mathbf{a}\cdot\mathbf{b})^2/2 + \mathbf{a}\cdot\mathbf{b} + 1 + \frac{1}{2} |\mathbf{a}\cdot \mathbf{b}|(\mathbf{a}\cdot\mathbf{b} + 2)\\
&=& (\mathbf{a}\cdot\mathbf{b})^2/2 + \mathbf{a}\cdot\mathbf{b} + 1 + \frac{1}{2} |\mathbf{a}\cdot\mathbf{b}|(\mathbf{a}\cdot\mathbf{b}) + |\mathbf{a}\cdot\mathbf{b}|.
\end{eqnarray*}
In the case of $\mathbf{a}\cdot\mathbf{b} \geq 0$,
\begin{eqnarray*}
\sigma_1^2 &=& |\mathbf{a}\cdot\mathbf{b}|^2/2 + |\mathbf{a}\cdot\mathbf{b}| + 1 + |\mathbf{a}\cdot\mathbf{b}|^2/2 + |\mathbf{a}\cdot \mathbf{b}|\\
&=&|\mathbf{a}\cdot\mathbf{b}|^2 + 2|\mathbf{a}\cdot\mathbf{b}| + 1\\
&\geq& 1\;
\end{eqnarray*}
In the case of $\mathbf{a}\cdot \mathbf{b} < 0$,
\begin{eqnarray*}
\sigma_1^2 &=& |\mathbf{a}\cdot\mathbf{b}|^2/2 - |\mathbf{a}\cdot\mathbf{b}| + 1 - |\mathbf{a}\cdot \mathbf{b}|^2/2 + |\mathbf{a}\cdot\mathbf{b}|\\
&=& 1\;.
\end{eqnarray*}
Combining these cases implies $\sigma_1^2 \geq 1$. Next, we show that $\sigma_2^2 \leq 1$:
\begin{eqnarray*}
\sigma_2^2 &=& x-y/2 \\
     &=& ||\mathbf{a}||^2||\mathbf{b}||^2/2 + \mathbf{a}\cdot \mathbf{b} + 1 - \frac{1}{2}|
|\mathbf{a}||||\mathbf{b}||\sqrt{||\mathbf{a}||^2||\mathbf{b}||^2 + 4 \mathbf{a}\cdot\mathbf{b} + 4}\\
&\leq&  ||\mathbf{a}||^2||\mathbf{b}||^2/2 + ||\mathbf{a}||||\mathbf{b}|| + 1 - \frac{1}{2}||\mathbf{a}||||\mathbf{b}||\sqrt{||\mathbf{a}||^2||\mathbf{b}||^2 + 4 ||\mathbf{a}||||\mathbf{b}|| + 4}\\
&=&  ||\mathbf{a}||^2||\mathbf{b}||^2/2 + ||\mathbf{a}||||\mathbf{b}|| + 1 - \frac{1}{2}||\mathbf{a}||||\mathbf{b}||(||\mathbf{a}||||\mathbf{b}||+2)\\
&=&  ||\mathbf{a}||^2||\mathbf{b}||^2/2 + ||\mathbf{a}||||\mathbf{b}|| + 1 - ||\mathbf{a}||^2||\mathbf{b}||^2/2 - ||\mathbf{a}||||\mathbf{b}||\\
&=& 1
\end{eqnarray*}
Then, since $\sigma_1^2 \geq 1$ and $\sigma_2^2 \leq 1$ , $\sigma_1 \geq 1$ and $\sigma_2 \leq 1$, so $\kappa = \sigma_1/\sigma_2$.

When $\alpha=1$, equation \ref{eq:general_kappa_def} reduces to 
\begin{equation}
\frac{\sigma_1}{\sigma_2} = \frac{\sqrt{3+2\cos\theta + \sqrt{5+4\cos\theta}}}{\sqrt{3+2\cos\theta - \sqrt{5+4\cos\theta}}}.
\end{equation}
\end{proof}

Note that when $\theta=\pi/2$ and $\alpha=1$, $\kappa=\frac{\sqrt{3+\sqrt{5}}}{\sqrt{3-\sqrt{5}}}=\frac{3+\sqrt{5}}{2}=1+\phi=\phi^2$, where $\phi=\frac{1+\sqrt{5}}{2}$ is the so-called golden ratio.
Further, this implies that when $\mathbf{a}$ and $\mathbf{b}$ are normalized random vectors, the condition number is approximately $\phi^2$ since for such random vectors, $\theta\approx0$ (see remark 3.25 here \cite{vershynin2018high}).
It is also worth noting that $\kappa\rightarrow\infty$ as $\theta\rightarrow \pi$.
This limiting behavior as $\theta\rightarrow\pi$ is not surprising, since $I+\mathbf{a}\mathbf{b}^T$ is singular when $\theta=\pi$.

The following theorem can be used to efficiently compute the condition number of low-rank modifications of unitary matrices. An algorithm for doing so is described after the proof of the theorem.
\begin{theorem}
    If $U,V$ are $N\times k$ complex matrices and $W$ is a unitary matrix whose first $K\le 2k$ columns span the column space of both $U$ and $V$, then the singular values of $I+UV^*$ are $\{1\}\cup S$, where $S$ is the set of singular values of $I+\tilde{U}\tilde{V}^*$, $U=\tilde{W}\tilde{U}$, $V=\tilde{W}\tilde{V}$, and $\tilde{W}$ is the $N\times K$ matrix whose columns are given by the first $K$ columns of $W$.
    \label{thm:rank_k_svd}
\end{theorem}
\begin{proof}
Let $W_i$ denote the i$^\mathrm{th}$ column of W.
\begin{eqnarray*}
    I + UV^*
    &=& I + (\tilde{W}\tilde{U})(\tilde{W}\tilde{V})^* \\
    &=& I + \tilde{W}\tilde{U}\tilde{V}^*\tilde{W}^* \\
    &=& I - \tilde{W}\tilde{W}^* + \tilde{W}(I + \tilde{U}\tilde{V}^*)\tilde{W}^* \\
    &=& \sum_{i=K+1}^N W_i W_i^* + \tilde{W}(I + \tilde{U}\tilde{V}^*)\tilde{W}^* \\
    &=& \sum_{i=K+1}^N W_i W_i^* + \tilde{W}A\Sigma B^* \tilde{W}^* \\
    &=& \sum_{i=K+1}^N W_i W_i^* + \sum_{i=1}^K (\tilde{W}A)_i\Sigma_i (B \tilde{W})^*_i
\end{eqnarray*}
where $A\Sigma B^*=I+\tilde{U}\tilde{V}^*$ is the singular value decomposition of $I+\tilde{U}\tilde{V}^*$, $(\tilde{W}A)_i$ is the i$^\mathrm{th}$ column of $(\tilde{W}A)$, $(\tilde{W}B)_i$ is the i$^\mathrm{th}$ column of $(\tilde{W}B)$, and $\Sigma_i$ is the i$^\mathrm{th}$ diagonal element of $\Sigma$.
The last equation is in the form of a singular value decomposition of $I+UV^*$, but we have not yet shown that $\{W_i\}_{i=K+1}^N\cup \{(\tilde{W}A)_i\}_{i=1}^K$ and $\{W_i\}_{i=K+1}^N\cup \{(\tilde{W}B)_i\}_{i=1}^K$ are orthonormal bases. We show this only for $\{W_i\}_{i=K+1}^N\cup \{(\tilde{W}A)_i\}_{i=1}^K$, since the case of $\{W_i\}_{i=K+1}^N\cup \{(\tilde{W}B)_i\}_{i=1}^K$ is essentially identical. There are three cases to consider:

Case 1: The two elements are both in $\{W_i\}_{i=K+1}^N$.
In this case, $W_i^*W_j=\delta_{ij}$ since $W$ is assumed to be unitary.

Case 2: One element is in $\{W_i\}_{i=K+1}^N$ and the other is in $\{(\tilde{W}A)_i\}_{i=1}^K$. In this case, $(\tilde{W}A)_j$ is in the subspace spanned by $\{W_i\}_{i=1}^K$ by construction of $\tilde{W}$. $W_i$ is orthogonal to every element in $\{W_i\}_{i=1}^K$ since $i>K$ and $W$ is a unitary matrix. Hence, $W_i^*(\tilde{W}A)_j=0$.

Case 3: Both elements are in $\{(\tilde{W}A)_i\}_{i=1}^K$. In this case,
\begin{eqnarray*}
    (\tilde{W}A)_i^* (\tilde{W}A)_j
    &=&  (\tilde{W}A_i)^* (\tilde{W}A_j) \\
    &=& A_i^*\tilde{W}^* \tilde{W}A_j \\
    &=& A_i^* A_j \\
    &=& \delta_{ij}
\end{eqnarray*}
Therefore, $\{W_i\}_{i=K+1}^N\cup \{(\tilde{W}A)_i\}_{i=1}^K$ is an orthonormal basis, and similarly for $\{W_i\}_{i=K+1}^N\cup \{(\tilde{W}B)_i\}_{i=1}^K$.

We now conclude that
\begin{equation*}
    I + UV^* = \sum_{i=K+1}^N W_i W_i^* + \sum_{i=1}^K (\tilde{W}A)_i\Sigma_i (B \tilde{W})^*_i
\end{equation*}
is a singular value decomposition of $I+UV^*$ and the singular values of $I+UV^*$ are 1 (for the $\sum_{i=K+1}^N W_i W_i^*$ terms) and $\{\Sigma_i\}_{i=1}^K$, where $\{\Sigma_i\}_{i=1}^K$ are the singular values of $I+\tilde{U}\tilde{V}^*$, as desired.
\end{proof}

Theorem \ref{thm:rank_k_svd} suggests an $O(Nk^2)$ algorithm for computing the singular value decomposition of $I+UV^*$ when $k\ll N$:
\begin{enumerate}
    \item Obtain $\tilde{W}$ by computing a thin singular value decomposition of the matrix formed by horizontally concatenating $U$ and $V$.
    \item Solve $\tilde{W}\tilde{U}=U$ for $\tilde{U}$.
    \item Solve $\tilde{W}\tilde{V}=V$ for $\tilde{V}$.
    \item Obtain $\Sigma$ by computing the singular value decomposition of $I+\tilde{U}\tilde{V}^*$.
    \item The singular values of $I+UV^*$ are then given by the diagonal of $\Sigma$ concatenated with a vector of length $N-2k$ whose elements are all 1.
    \item Sort the singular values from the previous step, if desired.
\end{enumerate}
This algorithm can also be used to compute the condition number, since the condition number is the ratio of the largest singular value to the smallest singular value.

\section{Application of zero noise extrapolation}
\label{app:zne}

We use zero noise extrapolation with 0 and 1 unitary foldings applied to the right-hand side of equation \ref{eq:case1final} as a whole rather than to each inner product.

The first step in the process is to transpile the circuits used to compute each inner product for the hardware.
This produces a series of $M$ unitaries, $Q_i$, that the hardware can execute.
The circuit can then be thought of as applying $Q_MQ_{M-1}\ldots Q_2 Q_1$ to the state $\ket{0}$.
Running this circuit and using it to estimate an inner product corresponds to having 0 ``foldings'' of the circuit.
Once each of the inner products is estimated, we obtain an estimate of $\braket{z}{x}$ by plugging the inner products into the right hand side of equation \ref{eq:case1final}.
We call this estimate $E_1$ where the subscript $1$ indicates the noise level.

The second step is to take the same circuits that were transpiled in the first step and apply a ``folding'' to each unitary.
Executing the circuit with a folding corresponds to applying
$Q_MQ_M^*Q_M Q_{M-1}Q_{M-1}^*Q_{M-1}\\\ldots Q_2Q_2^*Q_2 Q_1Q_1^*Q_1$ to the state $\ket{0}$ where $Q^*$ is the inverse of $Q$.
These circuits are theoretically identical to the previous circuits, except three times as long, which amplifies the noise.
All gates were folded, except for Qiskit's \verb+SXGate+ (a single qubit gate given by the square root of the Pauli $X$ operator).
This gate was excluded from the folding process because the inverse of the \verb+SXGate+ is not part of the hardware's native gate set.
Neglecting the \verb+SXGate+ in the folding process is unlikely to impact the results significantly since the noise is expected to be dominated by the two qubit gates.
Once each of the inner products is estimated using the folded circuits, we obtain an estimate of $\braket{z}{x}$ by again plugging the inner products into the right hand side of equation \ref{eq:case1final}.
We call the estimate with this folded circuit $E_3$.

With $E_1$ and $E_3$ in hand, we estimate
\begin{equation}
    E_0 = E_1 + \frac{(E_1 - E_3)}{2}
\end{equation}
as the zero noise extrapolation estimate of $\braket{z}{x}$.
This comes from fitting a line through the points $(1, E_1)$ and $(3, E_3)$, where the first coordinate gives the noise level and the second coordinate gives the estimate of $\braket{z}{x}$ at that noise level.
Finally, the curve is extrapolated to the zero noise level via the linear fit and that is used as the zero noise extrapolation estimate.

\bibliographystyle{abbrv}
\bibliography{refs}

\end{document}